\documentclass[pra,
twocolumn,
amsmath,amssymb,
superscriptaddress]{revtex4-2}
\usepackage{CJKutf8}
\usepackage{amsmath,amssymb,graphicx,color}
\usepackage{bm}
\usepackage{bbm}
\usepackage[caption=false]{subfig}  
\usepackage[usexnames,dvipsnames]{xcolor}
\usepackage{mathrsfs}
\usepackage[mathscr]{euscript}
\usepackage[normalem]{ulem}
\usepackage{soul}
\usepackage[colorlinks=true,citecolor=Cerulean,linkcolor=RubineRed,urlcolor=Cerulean]{hyperref}
\usepackage{soul,xcolor}

\begin{document}
\title{Local Non-Hermitian Hamiltonian Formalism for Dissipative Fermionic Systems and Loss-Induced Population Increase in Fermi Superfluids}

\author{Teng Xiao} \email{tengxiaozju@zju.edu.cn}
\affiliation{School of Physics and Zhejiang Institute of Modern Physics, Zhejiang University, Hangzhou, Zhejiang 310027, China}

\author{Gentaro Watanabe}\email{gentaro@zju.edu.cn}
\affiliation{School of Physics and Zhejiang Institute of Modern Physics, Zhejiang University, Hangzhou, Zhejiang 310027, China}
\affiliation{Zhejiang Province Key Laboratory of Quantum Technology and Device, Zhejiang University, Hangzhou, Zhejiang 310027, China}

\date{\today}
\begin{abstract}
  We examine a standard scheme to obtain the non-Hermitian Hamiltonian (NHH) from the Lindblad master equation by neglecting its jump term, and propose an alternative approach to address the limitations of the former. It is shown that the NHH obtained by the conventional scheme fails to provide a good approximation for fermionic many-body systems, even on short timescales. To resolve this issue, we present a framework called the local NHH formalism, which describes the loss process in each individual mode locally. This formalism is applicable to general dissipative fermionic systems and remains consistent with the underlying Lindblad master equation at the level of the equations of motion of local observables. The local NHH formalism also provides a convenient framework for studying non-Hermitian physics in dissipative fermionic systems, especially for spectral analysis, compared to the Lindblad master equation. As an illustration, we consider a fermionic superfluid subjected to one-body loss and find the population increase induced by the loss, resulting from the locking of the relative phase between the pairing gap and the anomalous field. The conventional NHH fails to capture these unique phenomena.
\end{abstract}

\maketitle

\noindent{\bf INTRODUCTION}\smallskip \\
Recent development of realizing many-body open quantum systems in various platforms with high controllability such as ultracold quantum gases~\cite{Muller2012review,Schafer2020review}, exciton-polaritons in semiconductor microcavities~\cite{Byrnes2014,Schneider2017,Bloch2022,Kavokin2022review}, and superconducting circuits~\cite{Houck2012review,Ma2021review} has opened a doorway to a new paradigm of non-Hermitian physics~\cite{Ashida2020review}. There, non-Hermiticity of the system caused by the interaction with the environment is utilized as a resource rather than things to be avoided, and rich phenomena that are inaccessible in closed Hermitian systems can be studied. Furthermore, exciting applications of non-Hermitian nature of open quantum systems to quantum technologies are in view. Along these lines, dissipative phase transitions~\cite{PhysRevX.7.011016,Fink2017,PhysRevLett.118.247402,PhysRevA.98.042118,PhysRevA.97.013825,PhysRevLett.129.087001}, measurement-driven phase transitions~\cite{Skinner2019jul,Li2018nov,Chan2019jun,Li2019oct,Lavasani2021mar,Noel2022jul,Koh2023sep}, and dissipative state preparation~\cite{Muller2012review,Diehl2008,PhysRevA.78.042307,verstraete2009quantum,PhysRevLett.105.227001,PhysRevLett.106.090502,Yi_2012,PhysRevA.85.023604,PhysRevA.89.013620,PhysRevLett.117.040501}, etc. are recently under active research.

Non-Hermitian Hamiltonian (NHH) formalism is a powerful effective theory for open quantum systems~\cite{moiseyev2011non,Ashida2020review}. Especially, it provides much simpler description of many-body systems compared to the other standard formalisms for open quantum systems~\cite{10.1093/acprof:oso/9780199213900.001.0001,wiseman_milburn_2009,jacobs_2014,doi:10.1080/00018732.2014.933502,RevModPhys.93.015008,Benatti2003text,Sieberer_2016}. The effective NHH description has made a great success for open bosonic systems such as photonic/optical systems~\cite{Feng2017review,ElGanainy2018review,  Guo2009aug,Ruter2010jan,Zeuner2015jul,Pan2018apr,Zhao2019sep,Xiao2020mar,Xia2021apr,Xiao2021jun,Wang2021jul,Lin2022sep,   Leykam2017jan,Lau2018oct,Zhou2019feb,Longhi2019jun,Pan2021may,Longhi2022aug}, atomic Bose–Einstein condensates (BECs)~\cite{Barontini2013jan,Labouvie2016jun,Tomita2017dec,Li2019feb,Lapp2019apr,Gou2020feb,Chen2021may,Liang2022aug,   Xu2017jan,Hamazaki2019aug,Li2020jun}, etc.
As a matter of fact, almost all the recent
major experimental achievements in the active study of the non-Hermitian quantum physics of the $\mathcal{PT}$-symmetry~\cite{Bender1998jun,Bender2007review,Bender2015review,  Feng2017review,ElGanainy2018review,Guo2009aug,Ruter2010jan,Regensburger2012aug,Hodaei2014nov,Li2019feb,Chen2021may} and topological phenomena under dissipation~\cite{Bergholtz2021review,Okuma2023review,Rudner2009feb,Esaki2011nov,   Zeuner2015jul,Pan2018apr,Zhao2019sep,Gou2020feb,Xiao2020mar,Xia2021apr,Xiao2021jun,Wang2021jul,Liang2022aug,Lin2022sep} are for the open bosonic systems
(see also Refs.~\cite{Ashida2020review,   Hatano1996jul,   Zhu2014jun,   Lee2014oct,   Lee2016apr,MartinezAlvarez2018mar,Xiong2018mar,Kunst2018jul,Lieu2018jan,Shen2018apr,Yao2018aug,Gong2018sep,MartinezAlvarez2018oct,Liu2019feb,Ghatak2019apr,Kawabata2019oct,Song2019dec,Okuma2020feb,Borgnia2020feb,   Hanai2019may} for other major theoretical developments of the related topics).

Spurred by the success of the NHH formalism for bosonic systems and recent experimental development of cold fermionic atoms under dissipation~\cite{Sponselee2018sep,Honda2023feb,Ren2022,zhao2023twodimensionalnonhermitianskineffect}, NHH for open fermionic systems becomes active area of research~\cite{Nakagawa2018nov,Ghatak2018jan,Yamamoto2019sep,SongPRL2019,ZHOU2019257,Nakagawa2020apr,Mu2020aug,Zhou2020oct,Liu2020dec,Iskin2021jan,Kanazawa2021,Zhou2021dec,Wang2022nov,Ding2022jan,Li2023feb,Tajima2023mar,Han2023jun,Takemori2023arxiv} (see also Refs.~\cite{Kantian2009dec,Foss-Feig2012dec,Yamamoto2021jul,Rosso2021nov,Li2022aug,Rosso2023jan,Mazza2023may,Dai2023sep} for other developments of the research in open fermionic many-body systems). NHH formalism for fermionic systems can be even more important compared to the bosonic counterpart because of their ubiquity in materials. However, almost all existing studies of the NHH formalism for fermionic systems are based on the NHH whose validity remains uncertain (see, e.g., Ref.~\cite{PhysRevA.108.032214}).
While some of these NHHs may be valid for engineered post-selection dynamics, addressing the natural evolution of the system without post-selection should be the primary concern.

In this paper, we take a significant step forward by proposing a new formalism, referred to as the local NHH formalism, which offers a suitable NHH description for purely dissipative fermionic systems in general.
When the underlying physics is well described by the Gorini-Kossakowski-Sudarshan-Lindblad master equation (hereafter, Lindblad master equation for short), it is believed that the NHH obtained by neglecting the jump term gives a good approximation in a sufficiently short timescale characterized by the inverse of the dissipation rate~{\cite{Ashida2020review,Yamamoto2019sep,PhysRevA.79.023614,PhysRevB.102.035153,PhysRevA.108.032214}.} Many previous studies of the NHH for open fermionic systems are based on this paradigm~\cite{Kantian2009dec,Nakagawa2018nov,Yamamoto2019sep,Nakagawa2020apr,Zhou2020oct,Liu2020dec,Iskin2021jan,Wang2022nov,Ding2022jan,Tajima2023mar,Han2023jun}. In the present work, we demonstrate that the NHH obtained by this scheme does not give a good approximation even in a sufficiently short timescale for dissipative fermionic systems. This crucial problem will be solved by our local NHH formalism, which provides the NHH consistent with the underlying Lindblad master equation at the level of equations of motion of observables considered. For illustration, by taking superfluid Fermi gases subject to one-body loss, we apply the local NHH formalism and find a unique phenomenon: an enhancement of the population induced by the loss, which the NHH obtained by the conventional scheme fails to reproduce even in a short timescale. The local NHH formalism proposed in our work provides a convenient pathway for further theoretical study of the non-Hermitian physics of dissipative fermionic systems compared to the original Lindblad master equation. As we demonstrate in this paper, spectral analysis of the local NHHs allows us to access the information directly related to the observables of interest, with much lower computational cost compared to the spectral analysis of the Liouvillian.
We set $\hbar=1$ throughout the paper.

\bigskip
\noindent{\bf RESULTS}\\
{\sf Model} \smallskip\\
By way of illustration, we consider a homogeneous two-component superfluid Fermi gas in three dimensions described by the Bardeen-Cooper-Schrieffer (BCS) mean-field Hamiltonian:
\begin{equation}\label{m2}
	\hat{H}_{\text{BCS}}= \sum_{\boldsymbol{k}\sigma}\epsilon_{\boldsymbol{k}}\hat{c}^{\dagger}_{\boldsymbol{k}\sigma}\hat{c}_{\boldsymbol{k}\sigma}
	-\left(\Delta\sum_{\boldsymbol{k}}\hat{c}^{\dagger}_{\boldsymbol{k}\uparrow}\hat{c}^{\dagger}_{-\boldsymbol{k}\downarrow}+\text{h.c.}\right),
\end{equation}
where $\epsilon_{\boldsymbol{k}}\equiv {\boldsymbol{k}^2}/{2m}$,
$m$ is the mass of the particles, and $\hat{c}^{\dagger}_{\boldsymbol{k}\sigma}$ ($\hat{c}_{\boldsymbol{k}\sigma}$) is the creation (annihilation) operator for fermions with momentum $\boldsymbol{k}$ and the spin projection $\sigma=\uparrow$, $\downarrow$. Pairing gap $\Delta$ is defined as $\Delta=-\frac{g}{V}\sum_{\boldsymbol{k}}\langle\hat{c}_{-\boldsymbol{k}\downarrow}\hat{c}_{\boldsymbol{k}\uparrow}\rangle$, where $\langle\cdots\rangle$ denotes an expectation value, $g$ is the coupling constant of the effective contact interaction, and $V$ is the volume of the system. All the above summations of $\boldsymbol{k}$ are taken up to the cutoff wavenumber $k_c$, and the coupling constant $g$ is renormalized as: $({k_F^3}/{8\pi E_F})({1}/{k_F a_s})={g}^{-1}+{V}^{-1}\sum_{\boldsymbol{k}(|\boldsymbol{k}|<k_c)}{1}/{2\epsilon_{\boldsymbol{k}}}$, where $E_F={k_F^2}/{2m}=\omega_F$ is the Fermi energy of a non-interacting Fermi gas with the same density, $k_F$ ($\omega_F$) is the corresponding Fermi wave vector (angular frequency), and $a_s$ is the $s$-wave scattering length. The ground state of $\hat{H}_{\text{BCS}}$ is $|\Psi_{\text{BCS}}\rangle=\prod_{\boldsymbol{k}}(u_{\boldsymbol{k}}+v_{\boldsymbol{k}}\hat{c}^{\dagger}_{\boldsymbol{k}\uparrow}\hat{c}^{\dagger}_{-\boldsymbol{k}\downarrow})|0\rangle$ with $|0\rangle$ being the vacuum state, $u_{\boldsymbol{k}}$ and $v_{\boldsymbol{k}}$ being the quasiparticle amplitudes at $\boldsymbol{k}$ satisfying $|u_{\boldsymbol{k}}|^2=\frac{1}{2}\left(1+{\xi_{\boldsymbol{k}}}/{E_{\boldsymbol{k}}}\right)$ and $|v_{\boldsymbol{k}}|^2=\frac{1}{2}\left(1-{\xi_{\boldsymbol{k}}}/{E_{\boldsymbol{k}}}\right)$. Here, $E_{\boldsymbol{k}}=\sqrt{(\epsilon_{\boldsymbol{k}}-\mu)^2+|\Delta|^2}$ is the BCS quasiparticle energy, $\xi_{\boldsymbol{k}}\equiv\epsilon_{\boldsymbol{k}}-\mu$, and $\mu$ is the chemical potential.

To be concrete, we consider one-body loss (extension for general loss processes such as two-body loss is straightforward) described by the following quantum stochastic master equation \cite{Ashida2020review} (see also \cite{wiseman_milburn_2009}) for density matrix $\hat{\rho}_c$ conditioned by a noise realization:
\begin{equation}\label{smeq}
	\begin{aligned}
		\dot{\hat{\rho}}_c =& -{i}[{\hat{H}_{\text{eff}}}\hat{\rho}_c-\hat{\rho}_c\hat{H}_{\text{eff}}^\dagger]+\sum_{\boldsymbol{k}\sigma}{\Gamma_{\boldsymbol{k}\sigma}}\langle\hat{c}^{\dagger}_{\boldsymbol{k}\sigma}\hat{c}_{\boldsymbol{k}\sigma}\rangle\hat{\rho}_c\\&+\sum_{\boldsymbol{k}\sigma}\left(\frac{\hat{c}_{\boldsymbol{k}\sigma}\hat{\rho}_c\hat{c}^{\dagger}_{\boldsymbol{k}\sigma}}{\langle\hat{c}^{\dagger}_{\boldsymbol{k}\sigma}\hat{c}_{\boldsymbol{k}\sigma}\rangle}-\hat{\rho}_c\right)\frac{\text{d}M_{\boldsymbol{k}\sigma}}{\text{d}t},
	\end{aligned}
\end{equation}
where $\hat{H}_{\text{eff}}=\hat{H}_{\text{BCS}}- i \sum_{\boldsymbol{k}\sigma}\frac{\Gamma_{\boldsymbol{k}\sigma}}{2}\hat{c}^{\dagger}_{\boldsymbol{k}\sigma}\hat{c}_{\boldsymbol{k}\sigma}$ is an effective NHH, $\Gamma_{\boldsymbol{k}\sigma}$ ($\geq 0$) is the loss rate for momentum $\boldsymbol{k}$ and spin $\sigma$, and the dot denotes the time derivative. ${\text{d}M_{\boldsymbol{k}\sigma}}$ is a random variable satisfying ${\text{d}M_{\boldsymbol{k}\sigma}}{\text{d}M_{\boldsymbol{k}'\sigma'}}=\delta_{\boldsymbol{k}\boldsymbol{k}'}\delta_{\sigma\sigma'}{\text{d}M_{\boldsymbol{k}\sigma}}$ whose mean value is $\mathcal{E}[{\text{d}M_{\boldsymbol{k}\sigma}}]={\Gamma_{\boldsymbol{k}\sigma}}\langle\hat{c}^{\dagger}_{\boldsymbol{k}\sigma}\hat{c}_{\boldsymbol{k}\sigma}\rangle{\text{d}t}$, where $\mathcal{E}[\cdot]$ denotes an ensemble average over the noise realizations. Note that Eq.~(\ref{smeq}) preserves the trace of $\hat{\rho}_c$: Namely, the change of the trace due to the non-Hermitian terms in $\hat{H}_{\text{eff}}$ is cancelled by the second term in the r.h.s. of Eq.~(\ref{smeq}) \cite{Zloshchastiev2014,Zloshchastiev2024}. The second line of Eq.~(\ref{smeq}) describes the discontinuous state change due to the particle loss in mode $(\boldsymbol{k},\sigma)$ under continuous measurement by the environment (or an experimental device) \cite{Ashida2020review}. The presence of the random variable $\text{d}M_{\boldsymbol{k}\sigma}$ in this term results from the probabilistic nature of the quantum measurement (see, e.g., Refs.~\cite{Ashida2020review,wiseman_milburn_2009,jacobs_2014} for more details about the stochastic master equation).  

By taking the ensemble average ($\hat{\rho}=\mathcal{E}[\hat{\rho}_c]$), Eq.\,\eqref{smeq} is reduced to the following Lindblad master equation:
\begin{equation}\label{lmeq}
	\begin{aligned}
	\dot{\hat{\rho}} = -{i}[{\hat{H}_{\text{BCS}}},\hat{\rho}] + \sum_{\boldsymbol{k}\sigma}\Gamma_{\boldsymbol{k}\sigma}\mathcal{D}[\hat{c}_{\boldsymbol{k}\sigma}]\hat{\rho} 
	\end{aligned}
\end{equation}
with the dissipator $\mathcal{D}[\hat{L}]\hat{\rho}\equiv\frac{1}{2}(2\hat{L}\hat{\rho}\hat{L}^{\dagger}-\hat{L}^{\dagger}\hat{L}\hat{\rho}-\hat{\rho}\hat{L}^{\dagger}\hat{L})$. The term $\hat{L}\hat{\rho}\hat{L}^{\dagger}$ in $\mathcal{D}[\hat{L}]\hat{\rho}$ is called the quantum jump term originated from the first term in the second line of Eq.~(\ref{smeq}). 
The effective NHH $\hat{H}_{\text{eff}}$ is obtained by neglecting the jump term as commonly employed in existing works \cite{doi:10.1080/00018732.2014.933502,Ashida2020review,Nakagawa2018nov,Yamamoto2019sep,Nakagawa2020apr,Zhou2020oct,Liu2020dec,Iskin2021jan,Wang2022nov,Ding2022jan,Tajima2023mar,Han2023jun,RevModPhys.70.101,PhysRevA.100.062131,PhysRevA.101.062112} which describes the time evolution of the system by the first line of Eq.\,\eqref{smeq}. Since the second line of Eq.\,\eqref{smeq} vanishes for the common eigenstate of all the Lindblad operators $\hat{c}_{\boldsymbol{k}\sigma}$'s, one can readily see that the NHH $\hat{H}_{\text{eff}}$ is valid for such a state, e.g., a Bose-Einstein condensate which is well-approximated by a coherent state. However, it is not the case for Fermi superfluids.

With this Lindblad master equation, we consider EOMs of the expectation value of the population of state $(\boldsymbol{k},\sigma)$, $n_{\boldsymbol{k}\sigma}\equiv \langle\hat{n}_{\boldsymbol{k}\sigma}\rangle\equiv \langle\hat{c}^{\dagger}_{\boldsymbol{k}\sigma}\hat{c}_{\boldsymbol{k}\sigma}\rangle$, and the anomalous average value of state $\boldsymbol{k}$, ${\nu}_{\boldsymbol{k}}\equiv \langle\hat{{\nu}}_{\boldsymbol{k}}\rangle\equiv\langle\hat{c}_{-\boldsymbol{k}\downarrow}\hat{c}_{\boldsymbol{k}\uparrow}\rangle$ (${\nu}^{*}_{\boldsymbol{k}}\equiv\langle\hat{{\nu}}^{\dagger}_{\boldsymbol{k}}\rangle\equiv\langle\hat{c}^{\dagger}_{\boldsymbol{k}\uparrow}\hat{c}^{\dagger}_{-\boldsymbol{k}\downarrow}\rangle$).
By substituting Eq.\,(\ref{lmeq}) into $\dot{n}_{\boldsymbol{k}\sigma}=\text{Tr}\left[\dot{\hat{\rho}}\hat{n}_{\boldsymbol{k}\sigma}\right]$ and $\dot{\nu}_{\boldsymbol{k}}=\text{Tr}\left[\dot{\hat{\rho}}\hat{{\nu}}_{\boldsymbol{k}}\right]$, and taking the transformations of ${\nu}_{\boldsymbol{k}}\rightarrow{\nu}_{\boldsymbol{k}}e^{-2i\mu t}$ and $\Delta\rightarrow\Delta e^{-2i\mu t}$, we get a set of closed coupled equations:
\begin{subequations}
\begin{align}
		\dot{n}_{\boldsymbol{k}\uparrow(-\boldsymbol{k}\downarrow)}=&\, {i}(\Delta{\nu}^{*}_{\boldsymbol{k}}-\Delta^*{\nu}_{\boldsymbol{k}}) -\Gamma_{\boldsymbol{k}\uparrow(-\boldsymbol{k}\downarrow)}n_{\boldsymbol{k}\uparrow(-\boldsymbol{k}\downarrow)},\label{eoma}\\		\dot{\nu}_{\boldsymbol{k}}=&\,-{i}2(\epsilon_{\boldsymbol{k}}-\mu)
		\nu_{\boldsymbol{k}} +{i}\Delta(1-n_{\boldsymbol{k}\uparrow}-n_{-\boldsymbol{k}\downarrow})\notag\\&\, -\frac{1}{2}(\Gamma_{\boldsymbol{k}\uparrow}+\Gamma_{-\boldsymbol{k}\downarrow})\nu_{\boldsymbol{k}}.\label{eomb}
\end{align}
\end{subequations}

For the evolution of the short timescale $t\ll\Gamma^{-1}$ starting from $\hat{\rho}=|\Psi_{\text{BCS}}\rangle\langle\Psi_{\text{BCS}}|$, the EOMs based on the NHH $\hat{H}_{\text{eff}}$, which are obtained from the first line of Eq.~(\ref{smeq}) [the second term of the r.h.s. is required for the trace conservation of $\hat{\rho}_c$], read:
\begin{subequations}
	\begin{align}
		\dot{n}_{\boldsymbol{k}\uparrow({-\boldsymbol{k}\downarrow})}&(0)= \big(\text{r.h.s. of Eq.\,\eqref{eoma}}\big)\notag\\&
		+|v_{\boldsymbol{k}}|^2\left[\Gamma_{-\boldsymbol{k}\downarrow({\boldsymbol{k}\uparrow})}(|v_{\boldsymbol{k}}|^2-1)
		+\Gamma_{\boldsymbol{k}\uparrow({-\boldsymbol{k}\downarrow})}|v_{\boldsymbol{k}}|^2\right],\label{eoma'}\\
		\dot{\nu}_{\boldsymbol{k}}(0)=&\, \big(\text{r.h.s. of Eq.\,\eqref{eomb}}\big)\notag\\&\, +u^*_{\boldsymbol{k}}v_{\boldsymbol{k}}(\Gamma_{-\boldsymbol{k}\downarrow}|v_{\boldsymbol{k}}|^2+\Gamma_{\boldsymbol{k}\uparrow}|v_{\boldsymbol{k}}|^2)\label{eomb'}
	\end{align}
\end{subequations} 
with $n_{\boldsymbol{k}\sigma}(0)=|v_{\boldsymbol{k}}|^2$, ${\nu}_{\boldsymbol{k}}(0)=u^*_{\boldsymbol{k}}v_{\boldsymbol{k}}$, and $u_{\boldsymbol{k}}$ and $v_{\boldsymbol{k}}$ being their equilibrium values. Note that the EOMs \eqref{eoma'} and \eqref{eomb'} based on $\hat{H}_{\text{eff}}$ are not consistent with the corresponding EOMs \eqref{eoma} and \eqref{eomb} obtained from the original master equation. 
For example, let us discuss the difference between Eqs.\,\eqref{eoma} and \eqref{eoma'} when $\Gamma_{\boldsymbol{k}\sigma}$'s for all the states $(\boldsymbol{k},\sigma)$ are the same. The first term in the square brackets of Eq.\,\eqref{eoma'}, which comes from the pairing interaction, vanishes in the BCS limit. On the other hand, the second term, which originates from the difference between the initial BCS state and the eigenstate of the Lindblad operator, becomes negligible in the BEC limit. Therefore, these two extra terms do not vanish simultaneously, and they give the contributions comparable to the loss term in Eq.\,\eqref{eoma} in general.

The loss described by the Lindblad master equation is an incoherent process because the quantum jump yields a statistical mixture of pure states.
However, the conventional NHH formalism without the quantum jump is a coherent loss description.
Because of this artificial treatment of the dissipation by neglecting the quantum jump term, we get those extra terms in Eqs.\,\eqref{eoma'} and \eqref{eomb'}. In short, the NHH scheme of neglecting the jump term can be invalid even in a short timescale in fermionic many-body systems. For the local observables of each mode like $n_{\boldsymbol{k}\sigma}$ and $\nu_{\boldsymbol{k}}$, this timescale means $t \ll \Gamma^{-1}$.

\begin{figure}[!tbp]
	\centering \includegraphics[width=1\linewidth]{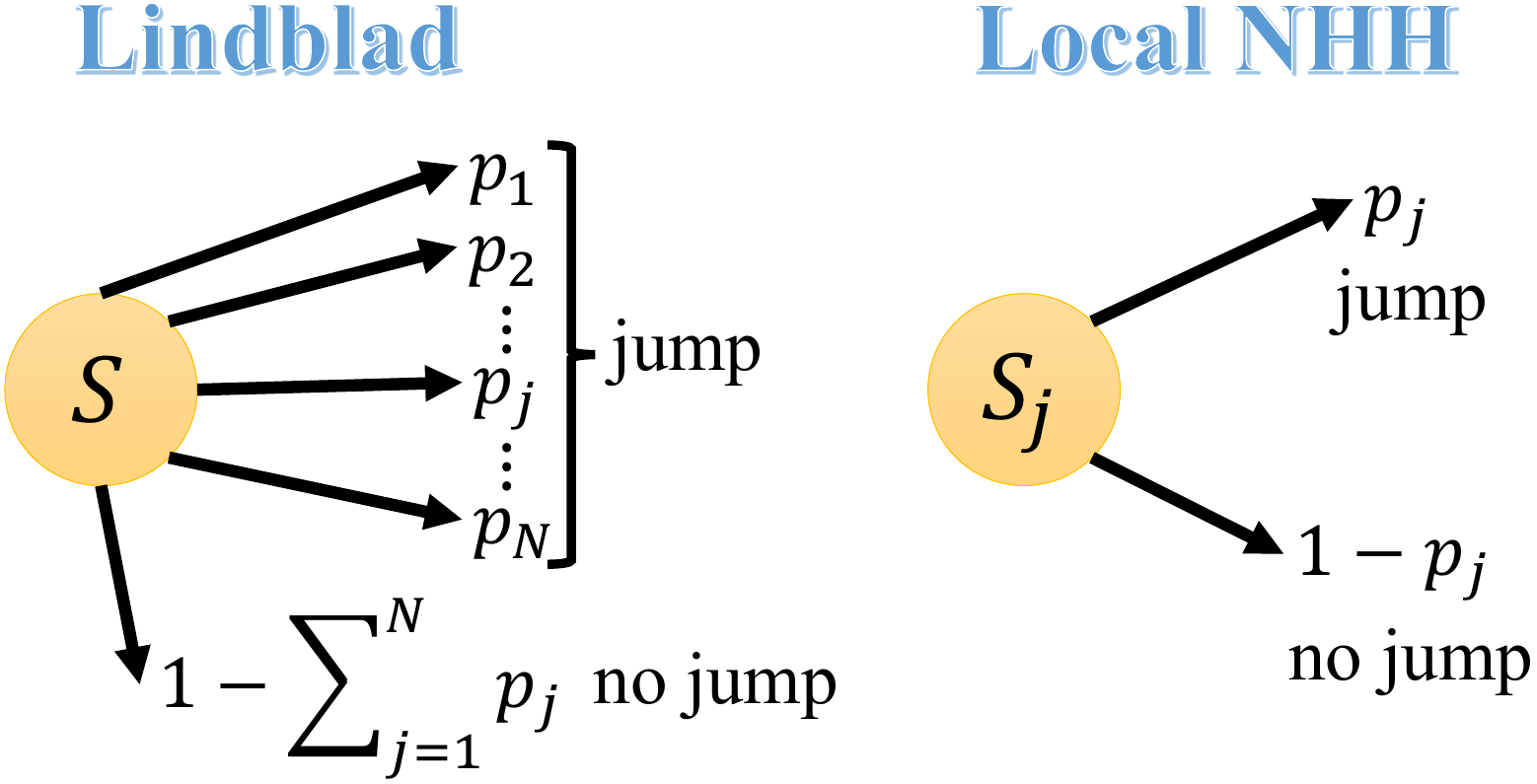} \caption{{\bf Illustration of the difference between the Lindblad master equation and the local NHH.} For the system $S$ with $N$ modes, statistical mixture of all the possible jumps and jump-free case can be considered in the case of the Lindblad master equation. However, we need to focus only on the jump-free process of each individual mode $j$ in the NHH case because of its inability to describe the mixed state. In this figure, we consider a local operator associated with a single mode $j$ for simplicity.}
	\label{FIG_illust}
\end{figure}

\bigskip
\noindent{\sf Local NHH formalism}\smallskip\\
Here, we introduce the local NHH formalism applicable to dissipative fermionic many-body systems. We consider a general fermionic system with $N$ modes whose Hamiltonian is $\hat{H}$. (While we specifically consider the BCS Hamiltonian in later discussions, the local NHH formalism is applicable to general fermionic Hamiltonians, including non-quadratic forms and cases with mode coupling.) This system is under local dissipation for each mode $j$ described by the Lindblad operator $\hat{L}_j$ for particle loss processes (e.g., one-body loss, two-body loss, etc.) with the dissipation rate $\Gamma_j$. (In this paper, the term ``local'' refers to locality with respect to mode, regardless of the choice of basis.) In deriving the Lindblad master equation, statistical mixture of all the possible jump processes and a jump-free case are considered (left panel of Fig.\,\ref{FIG_illust}). With the probability $p_j$ of a jump by $\hat{L}_j$ in $\text{d}t$, the probability of the jump-free process is $1- \sum_{j=1}^{N} p_j$ and its corresponding NHH is $\hat{H}_{\text{eff}}=\hat{H}- i\sum_j \frac{\Gamma_j}{2} \hat{L}_j^\dagger \hat{L}_j$.

Next, we consider a local quantity consisting of a single term (``single-term local quantity'' in short), $\hat{O}_j$, for mode $j$ (e.g., $\hat{n}_{\boldsymbol{k}\sigma}$ of state ($\boldsymbol{k},\sigma$), $\hat{{\nu}}_{\boldsymbol{k}}$ of state $\boldsymbol{k}$, etc.) in normal order, and study its time evolution within the NHH formalism.
For example, $\hat{n}_{\boldsymbol{k}\sigma}$ and $\hat{{\nu}}_{\boldsymbol{k}} = \hat{c}_{-\boldsymbol{k} \downarrow} \hat{c}_{\boldsymbol{k} \uparrow}$ are single-term local quantities, whereas $\hat{n}_{\boldsymbol{k} \uparrow} + \hat{n}_{-\boldsymbol{k} \downarrow}$ is not. It should also be noted that ``mode $j$'' does not necessarily correspond to a single mode of the system but may include two or more modes, depending on the specific form of $\hat{O}_j$. For instance, in the case of $\hat{O}_j=\hat{\nu}_{\boldsymbol{k}}$, ``mode $j$'' represents a set of two modes of $({\boldsymbol{k},\uparrow})$ and $({-\boldsymbol{k},\downarrow})$.

Since the Schr\"{o}dinger equation (or the von Neumann equation) based on the NHH cannot describe the statistical mixture, we must focus only on the dissipation process for each individual mode $j$ without being concerned about the other modes (right panel of Fig.\,\ref{FIG_illust}). Then the corresponding local NHH is\begin{equation}\label{nhh}
	\begin{aligned}
		\hat{H}_{\text{eff},j} = \hat{H}- i\frac{\Gamma_j}{2} \hat{L}_j^\dagger \hat{L}_j,
	\end{aligned}
\end{equation} 
and the stochastic master equation is given by Eq.\,\eqref{smeq} with $\hat{H}_{\text{eff}}$ being replaced by $\hat{H}_{\text{eff},j}$ [Eq.\,\eqref{nhh}].
For $\hat{O}_j$ associated with a set of multiple modes $\{l\}_j$, such as $\hat{{\nu}}_{\boldsymbol{k}}$, the local NHH is given by
\begin{equation}\label{nhh2}
	\begin{aligned}
		\hat{H}_{\text{eff},j} = \hat{H}- i\sum_{\{l\}_j}\frac{\Gamma_l}{2} \hat{L}_{l}^\dagger \hat{L}_{l}
	\end{aligned}
\end{equation}
[see Methods for detailed derivation of Eqs.~(\ref{nhh}), (\ref{nhh2}), and (\ref{meq})].
After taking the ensemble average with this stochastic master equation, the second term in the r.h.s. of Eq.\,\eqref{smeq} and the second term in the second line of Eq.\,\eqref{smeq} cancel.
The contribution from the jump term in $\langle\hat{O}_j\rangle$ vanishes for normally ordered operator $\hat{O}_j$ in the case of fermionic systems because of the consecutive repetition of the annihilation or creation operators of the same mode. Although $\hat{O}_j$ is required to be normally ordered, this class of operators encompasses a wide range of physically relevant local quantities.
Thereupon, the EOM of ${O}_j \equiv \langle\hat{O}_j\rangle$ becomes:
\begin{equation}\label{meq}
  \begin{aligned}
    \dot{O}_j = i \langle \hat{H}^{\dagger}_{\text{eff}, j} \hat{O}_j - \hat{O}_j \hat{H}_{\text{eff}, j} \rangle.
  \end{aligned}
\end{equation}
with $\hat{H}_{\text{eff}, j}$ given by Eqs.~(\ref{nhh}) or (\ref{nhh2}).
For the system considered in the previous section, an appropriate NHH to describe the time evolution of ${n}_{\boldsymbol{k}\sigma}$ is $\hat{H}_{\text{eff},(\boldsymbol{k},\sigma)}=\hat{H}_{\text{BCS}}-i \frac{\Gamma_{\boldsymbol{k}\sigma}}{2}\hat{c}^{\dagger}_{\boldsymbol{k}\sigma}\hat{c}_{\boldsymbol{k}\sigma}$ with $j \rightarrow(\boldsymbol{k},\sigma)$ and $\hat{L}_j\rightarrow\hat{c}_{\boldsymbol{k}\sigma}$. 
(The local NHH to describe the time evolution of ${\nu}_{\boldsymbol{k}}$ is $\hat{H}_{\text{eff},(\boldsymbol{k})}=\hat{H}_{\text{BCS}}-i \frac{\Gamma_{\boldsymbol{k}\uparrow}}{2}\hat{c}^{\dagger}_{\boldsymbol{k}\uparrow}\hat{c}_{\boldsymbol{k}\uparrow}-i \frac{\Gamma_{-\boldsymbol{k}\downarrow}}{2}\hat{c}^{\dagger}_{-\boldsymbol{k}\downarrow}\hat{c}_{-\boldsymbol{k}\downarrow}$ with $\{l\}_j \rightarrow \{({\boldsymbol{k},\uparrow}),({-\boldsymbol{k},\downarrow})\}_{\boldsymbol{k}}$.)
With this $\hat{H}_{\text{eff},(\boldsymbol{k},\sigma)}$, Eq.~\eqref{meq} yields the correct EOM given by Eq.~\eqref{eoma}.
In other words, the local NHH formalism is an effective description that produces the same EOMs as those derived from the original Lindblad master equation.
While it yields the same EOMs, the local NHH formalism is not equivalent to the original Lindblad master equation. As discussed later and demonstrated in the Methods section, the local NHH formalism provides an effective description that offers a more convenient framework for further theoretical analysis of the system. Notably, unlike the conventional NHH obtained by neglecting the jump term, the local NHH formalism is applicable to arbitrary timescales.

According to the above argument, for systems with multiple dissipative modes, an appropriate NHH differs by the mode and physical quantity considered. Since the NHH formalism cannot describe the statistical mixture generated by the quantum jump, the whole system cannot be described by a single NHH. Instead, a set of NHHs, $\{\hat{H}_{\text{eff}, j}\}$, for each local process $\hat{L}_j$ must be prepared.
Only for the system with a single dissipative mode, $\hat{H}_{\text{eff}}$ becomes identical to $\hat{H}_{\text{eff}, j}$, so that $\hat{H}_{\text{eff}}$ becomes valid with Eq.~\eqref{meq}. In other words, the commonly employed scheme to obtain the NHH by neglecting the jump term of the Lindblad master equation is inappropriate except for the system with a single dissipative mode such as a photon field in a single-mode optical cavity and an atomic Bose-Einstein condensate under particle loss. Our scheme to construct a set of appropriate NHHs is applicable not only to the purely dissipative fermionic superfluids specifically considered in the present work, but also to the purely dissipative fermionic many-body systems in general.

\begin{figure}[!tbp]
  \centering \includegraphics[width=1\linewidth]{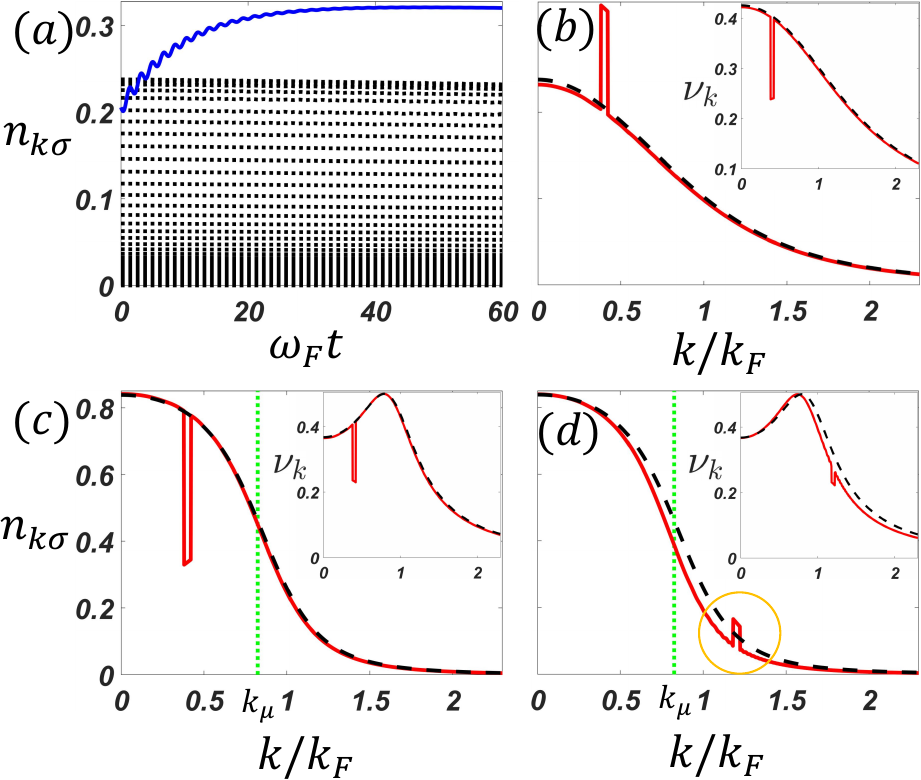}
  \caption{{\bf Dynamics of the population $n_{k\sigma}$ and loss-induced population increase in the quasi-steady state.} (a): Time evolution of $n_{k\sigma}$. The blue line is $n_{k\sigma}$ in the loss region. $n_{k\sigma}$'s in the region without the loss are shown by the black dotted lines. For clarity, we plot $n_{k\sigma}$ only for $k_{\text{cen}}$ in the loss region and for $k$'s at intervals of $2\delta_{{k}}$ in the loss-free region. (b), (c), and (d): Initial (black dashed line) and final (red solid line) profiles of $n_{k\sigma}$ in the time evolution for $\omega_Ft=60$. Same for the profiles of ${\nu}_{{k}}$ in the insets of (b), (c), and (d). $k_{\mu}$ is indicated by the vertical dotted line in (c) and (d). Here, ${k}_{\text{cen}}=0.4\,k_F \,[1.2\,k_F]$ in (a), (b), and (c) [(d)]. We set ${1}/{k_F a_s}=1$, $k_c = 15\,k_F$ in (a) and (b), and ${1}/{k_F a_s}=0$ and $k_c = 10\,k_F$ in (c) and (d).}
	\label{FIG_pop}
\end{figure}

\bigskip
\noindent{\sf Loss-induced population increase}\smallskip\\
Next, we demonstrate population increase induced by the particle loss in superfluid Fermi gases after long-time evolution. The particle loss is introduced in a narrow region in momentum space,  $|\boldsymbol{k}_{{\Gamma}}|\in\left[{k}_{\text{cen}}-\delta_k/2,{k}_{\text{cen}}+\delta_k/2 \right]$. Here, $\boldsymbol{k}_{{\Gamma}}$ denotes the momentum in this loss region, ${k}_{\text{cen}}$ is the center and $\delta_k$ is the width of this region. We set the dissipation rate is constant $\Gamma_{\boldsymbol{k}\sigma}=\Gamma$ in this region. The ground state $|\Psi_{\text{BCS}}\rangle$ is taken as the initial state. (As demonstrated in Methods, even if starting from a different initial state other than $|\Psi_{\text{BCS}}\rangle$, the main results discussed in this section can also be obtained.) In the following, we perform the analysis and numerical simulation based on the EOMs \eqref{eoma} and \eqref{eomb}, which are also obtained from the local NHH. Throughout the paper, we set $\delta_{{k}}=0.04\,k_F$ and $\Gamma =0.1\,\omega_F$ (momentum is discretized by $0.002\,k_F$ in the numerical simulation). Qualitative results do not change for different values of $\delta_k$ and $\Gamma$ as far as $\delta_k\,\Gamma$ is small enough so that $|\Delta|$ changes slowly over time.

In Fig.~\ref{FIG_pop}(a), we show the time evolution of $n_{k\sigma}$ for ${1}/{k_F a_s}=1$ (BEC regime). The results are the same for spin up and down because of the isotropy of the system. It is remarkable that the population $n_{k\sigma}$ in the loss region (blue solid line) increases and converges to a value larger than the initial one at $t\gtrsim\Gamma^{-1}=10/\omega_F$. Then the system becomes almost steady, which we call as a quasi-steady state. Although this quasi-steady state is not exactly steady and the total number of particles keeps on decreasing slowly due to the loss, it appears for a long time during which the system is practically unchanged.
In Fig.~\ref{FIG_pop}(b), the prominent enhancement of $n_{{k}\sigma}$ at $k_{\Gamma}$ in the final state (at $\omega_F t=60$; red solid line) can be observed. This loss-induced population increase is more prominent in the deep BEC side with stronger interaction. Although the population at $\boldsymbol{k}_{{\Gamma}}$ increases, ${\nu}_{{k}}$ still decreases as shown in the inset of Fig.~\ref{FIG_pop}(b). This is because the Cooper pairs of atoms with $\pm\boldsymbol{k}_{{\Gamma}}$ are broken due to the one-body loss, and the increased population in the loss region consists of particles without pairing. Whether $n_{\boldsymbol{k}_{\Gamma}\sigma}$ is increased or decreased from its initial value is determined by whether $k_{\text{cen}}$ is larger or smaller than $k_{\mu}\equiv\sqrt{2m\mu}$ as shown in Figs.~\ref{FIG_pop}(c) and \ref{FIG_pop}(d) for $k_{\text{cen}}<k_{\mu}$ and $k_{\text{cen}}>k_{\mu}$ at $1/k_Fa_s = 0$ (unitarity), respectively.

\begin{figure}[!tbp]
	\centering \includegraphics[width=1\linewidth]{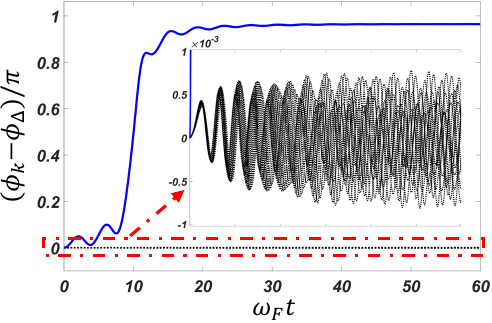} \caption{ {\bf Dissipation-induced phase locking.} Time evolution of the relative phase $(\phi_{k}-\phi_{\Delta})$ for various values of $k$. The blue line is the relative phase at $k_{\Gamma}$ in the loss region. $(\phi_{k}-\phi_{\Delta})$'s in the region without the loss are shown by the black dotted lines. The inset is the magnification of the red dash-dotted box. For clarity of the figure, we plot $(\phi_{k}-\phi_{\Delta})$ only for $k_{\text{cen}}$ (blue line) in the loss region and for $k$'s (dotted line) at intervals of $\delta_{k}$ in the loss-free region in $[{k}_{\text{cen}}-11\delta_{{k}}/2,{k}_{\text{cen}}+11\delta_{{k}}/2]$. The other parameters are same as Fig.\,\ref{FIG_pop}(c) (${1}/{k_F a_s}=0$, $k_c=10\,k_F$, $k_{\text{cen}}=0.4\,k_F$, $\delta_{{k}}=0.04\,k_F$, and $\Gamma =0.1\,\omega_F$).}
	\label{FIG_phase}
\end{figure}

To interpret the above phenomenon, it is important to note that $\Delta{\nu}^{*}_{\boldsymbol{k}}-\Delta^*{\nu}_{\boldsymbol{k}}$ is pure imaginary, and the first term of Eq.\,\eqref{eoma}, ${i}(\Delta{\nu}^{*}_{\boldsymbol{k}}-\Delta^*{\nu}_{\boldsymbol{k}})={2}|\Delta||{\nu}_{\boldsymbol{k}}|\sin(\phi_{\boldsymbol{k}}-\phi_{\Delta})$, can be regarded as an effective pump/loss depending on the relative phase $(\phi_{\boldsymbol{k}}-\phi_{\Delta})$ between $\Delta=|\Delta|e^{i\phi_{\Delta}}$ and ${\nu}_{\boldsymbol{k}}=|{\nu}_{\boldsymbol{k}}|e^{i\phi_{\boldsymbol{k}}}$.
As shown in Fig.~\ref{FIG_phase}, in the presence of loss,
the relative phases $(\phi_{\boldsymbol{k}}-\phi_{\Delta})$ for $\boldsymbol{k}$'s in the loss region (blue solid line) stay at a fixed nonzero constant value after long time ($t\gg\Gamma^{-1}$) which we call the dissipation-induced phase locking in the quasi-steady state. On the other hand, as the black dotted lines in Fig.~\ref{FIG_phase} show (clearer in the inset), the values of the relative phases $(\phi_{\boldsymbol{k}}-\phi_{\Delta})$'s in the loss-free region spread over time. Although the amplitude of its variation is small, its behavior is qualitatively different from the one in the loss region.
While Fig.~\ref{FIG_phase} shows the case of ${1}/{k_F a_s}=0$ with $\delta_{{k}}=0.04\,k_F$ and $\Gamma =0.1\,\omega_F$ as a concrete example, the dissipation-induced phase locking can be observed throughout the BCS-BEC crossover as far as $\delta_k\,\Gamma$ is small enough so that the system evolves into a quasi-steady state.

To get an approximate analytical quasi-steady solution in the loss region, we use the condition that $|\Delta|$ changes very slowly.
Thus, by setting the l.h.s. of Eqs.\,\eqref{eoma} and \eqref{eomb} equal to zero, we obtain (the superscript ``qs'' represents the quasi-steady solution): 
\begin{align}\label{qst}
		&n^{\text{qs}}_{\boldsymbol{k}_{{\Gamma}}\uparrow}=n^{\text{qs}}_{-\boldsymbol{k}_{{\Gamma}}\downarrow}=\frac{1}{2}\frac{|\Delta|^2}{|\Delta|^2+(\epsilon_{\boldsymbol{k}_{{\Gamma}}}-\mu)^2+{\Gamma^2}/{4}}\notag,\\
		&|\nu^{\text{qs}}_{\boldsymbol{k}_{{\Gamma}}}|=\frac{1}{2}\frac{|\Delta|\sqrt{(\epsilon_{\boldsymbol{k}_{{\Gamma}}}-\mu)^2+{\Gamma^2}/{4}}}{|\Delta|^2+(\epsilon_{\boldsymbol{k}_{{\Gamma}}}-\mu)^2+{\Gamma^2}/{4}},\\
		&(\phi_{\boldsymbol{k}_{{\Gamma}}}-\phi_{\Delta})^{\text{qs}}=\arccos\left(\frac{(\epsilon_{\boldsymbol{k}_{{\Gamma}}}-\mu)}{\sqrt{(\epsilon_{\boldsymbol{k}_{{\Gamma}}}-\mu)^2+{\Gamma^2}/{4}}}\right)\in[0,\pi]\notag.
\end{align}
This analytical quasi-steady solution is verified by the numerical calculation as demonstrated in Methods.
Since $(\phi_{\boldsymbol{k}_{{\Gamma}}}-\phi_{\Delta})^{\text{qs}}$ is in the range between $0$ and $\pi$, the term ${2}|\Delta||{\nu}_{\boldsymbol{k}_{{\Gamma}}}|\sin(\phi_{\boldsymbol{k}_{{\Gamma}}}-\phi_{\Delta})>0$ acts as an effective pump in the loss region for any value of $(\epsilon_{\boldsymbol{k}_{{\Gamma}}}-\mu)$. Due to the presence of $\Delta$, all the momentum states are coupled through $\Delta$. Therefore, if there is a sink at some $k_\Gamma$, a flow in the momentum space occurs. Depending on the value of $k_\Gamma$, the flux to $k_\Gamma$ can be bigger than the flux of the removed particles from $k_\Gamma$ resulting in the increase of the population at $k_\Gamma$. To demonstrate the non-trivial population increase in the loss region quantitatively, we calculate the difference ${\delta}n_{\boldsymbol{k}_{\Gamma}}$ between $n^{\text{qs}}_{\boldsymbol{k}_{{\Gamma}}\sigma}$ and its initial value, ${\delta}n_{\boldsymbol{k}_{\Gamma}} = n^{\text{qs}}_{\boldsymbol{k}_{\Gamma}\sigma}-n_{\boldsymbol{k}_{\Gamma}\sigma}(0)$,
and its ratio against  $n_{\boldsymbol{k}_{\Gamma}\sigma}(0)$:
\begin{equation}\label{dn}
	\begin{aligned}
	\frac{{\delta}n_{\boldsymbol{k}_{{\Gamma}}}}{n_{\boldsymbol{k}_{{\Gamma}}\sigma}(0)}=\frac{E_{\boldsymbol{k}_{{\Gamma}}}(\epsilon_{\boldsymbol{k}_{{\Gamma}}}-\mu)-{\Gamma^2}/{4}}{E^2_{\boldsymbol{k}_{{\Gamma}}}+{\Gamma^2}/{4}}.
	\end{aligned}
\end{equation}
From Eq.\,\eqref{dn}, the sign of ${\delta}n_{\boldsymbol{k}_{{\Gamma}}}$ depends on the chemical potential $\mu$ and loss rate $\Gamma$. When $\Gamma$ is small, ${\delta}n_{\boldsymbol{k}_{{\Gamma}}}$ is positive (negative) if the kinetic energy in the loss region $\epsilon_{\boldsymbol{k}_{{\Gamma}}}$ is bigger (smaller) than $\mu$. With increasing ${1}/{k_F a_s}$, $\mu$ decreases from positive to negative value monotonically. Thus the ratio given by Eq.\,\eqref{dn} becomes positive and further increases with increasing ${1}/{k_F a_s}$.

The loss-induced population increase found in our work is unique for fermionic superfluids. In this phenomenon, the presence of another field ${{\nu}}_{\boldsymbol{k}}$, which works as a source of particles, in addition to ${n}_{\boldsymbol{k}\sigma}$ is essential. This is in a striking difference from bosonic superfluids where the magnitude of the superfluid order parameter and the density field are essentially the same.

The typical timescale to develop the population increase is $\Gamma^{-1}$. As shown in Fig.~\ref{FIG_pop}(a), the population increase becomes already apparent at $t\gtrsim2\Gamma^{-1}$, which corresponds to $\gtrsim 200\,\mu$s in typical experiments with the Fermi temperature $T_F\equiv E_F/k_{B}=1\,\mu \text{K}$ ($k_{B}$ is the Boltzmann constant) when $\Gamma=0.1\,\omega_F$.  This timescale is easily accessible in the current experiments of superfluid atomic Fermi gases \cite{PhysRevLett.88.120405,PhysRevLett.98.030401,PhysRevLett.105.030402,PhysRevLett.108.215301} whose typical lifetime is a few seconds. After a long time-of-flight, the population profile in momentum space is detected clearly.  
A promising setup to realize the narrow loss region in the momentum space and observe the population increase is quasi one-dimensional ultracold Fermi gases. With a highly accurate frequency-locked laser, atoms with a given momentum will be kicked out from the system (see also \cite{PhysRevA.107.023321}). The mechanism to make the pairing gap into an effective particle source or sink using the state-selective loss should find interesting applications in the future quantum technology of, e.g., atomtronic circuits of fermionic superfluids~\cite{PhysRevLett.128.150401}.

\bigskip
\noindent{\bf DISCUSSION}\smallskip\\
The common scheme to obtain the NHH from the underlying Lindblad master equation by neglecting its jump term is invalid for fermionic systems (more precisely, multi-mode systems in general) even for short time scales, as it becomes inconsistent when the system is a statistical mixture in different modes. The local NHH formalism presented in our work resolves this issue by providing NHHs that are consistent with the underlying master equation at the level of equations of motion of local quantities.

Our formalism enables the study of the non-Hermitian physics of dissipative fermionic systems and suggests the need to revise previous results in the field.
Applying our formalism to fermionic superfluids under one-body loss, we find a unique non-Hermitian phenomenon where the loss enhances the population. This work serves as a stepping stone towards the NHH for general open quantum systems, and is expected to have broad applications in non-Hermitian physics of many-body open quantum systems.

Although the local NHH formalism yields the same equations of motion as the original master equation, it is based on an $N$-dimensional Hilbert space, while the original master equation uses an $N^2$-dimensional Liouville space. If we perform a spectral analysis of all the local NHHs, whose number is at most on the order of $N$, the total number of matrix elements involved is at most on the order of $N\times (N \times N)$. On the other hand, the Liouvillian superoperator matrix of the Lindblad master equation contains $N^2 \times N^2$ elements.
As a result, spectral analysis is significantly more tractable in the local NHH formalism. Furthermore, as demonstrated for a simple two-mode model in Methods (the section of ``Spectral analysis of the local NHH for a simple model''), the eigenspectrum of the local NHH carries the information of the dynamics of local observables, which is difficult to extract from the spectra of the conventional NHH or the Liouvillian of the original master equation.

It is also noteworthy that the dynamics of the local observables can exhibit interesting phenomena which can be observed in experiments. The population increase induced by the loss in the superfluid Fermi gases, found in this work, holds potential for future applications in quantum technologies.
A significant advantage of the local NHH formalism is its ability to provide convenient access to information that is otherwise hard to access through other formalisms. Further studies of the local NHH spectrum, including detailed comparisons with the corresponding Liouvillian spectrum and the spectrum of the conventional NHH for various models, are expected to offer valuable insights into the non-Hermitian physics of fermionic systems.

\bigskip
\noindent{\bf METHODS}\smallskip\\
Here, we show the details about the local NHH formalism and further discussions of the results presented in the previous sections. Specifically, we provide: 1) derivation of the local NHH formalism, 2) justification of the analytical quasi-steady solution given by Eq.~\eqref{qst}, 3) demonstration of the robustness of the loss-induced population increase, and 4) spectral analysis of the local NHH for a simple model.

\medskip
\noindent{\sf Derivation of the local NHH formalism}\smallskip\\
Here we provide the details about the derivation of Eqs.~\eqref{nhh} and \eqref{meq}. We consider a general fermionic system with the Hamiltonian $\hat{H}$ subject to the local dissipation for each mode $j$ described by the Lindblad operator $\hat{L}_j$ for particle loss processes, which can be described by the following stochastic master equation:
\begin{equation}\label{smeq2}
	\begin{aligned}
		\dot{\hat{\rho}}_c =& -{i}[{\hat{H}_{\text{eff}}}\hat{\rho}_c-\hat{\rho}_c\hat{H}_{\text{eff}}^\dagger]+\sum_{j}{\Gamma_{j}}\langle\hat{L}_j^{\dagger}\hat{L}_j\rangle\hat{\rho}_c\\&+\sum_{j}\left(\frac{\hat{L}_{j}\hat{\rho}_c\hat{L}^{\dagger}_{j}}{\langle\hat{L}_j^{\dagger}\hat{L}_j\rangle}-\hat{\rho}_c\right)\frac{\text{d}M_{j}}{\text{d}t}
	\end{aligned}
\end{equation}
with $\hat{H}_{\text{eff}} \equiv \hat{H}- i\sum_j \frac{\Gamma_j}{2} \hat{L}_j^\dagger \hat{L}_j$.
By taking an ensemble average over the noise realizations, $-\hat{\rho}_c\, \text{d}M_j/\text{d}t$ term in the second line becomes $-\hat{\rho}\, \Gamma_j\, \langle \hat{L}_j^\dagger \hat{L}_j \rangle$ and the term $\Gamma_j\, \langle \hat{L}_j^\dagger \hat{L}_j \rangle\, \hat{\rho}_c$ in the first line becomes $\Gamma_j\, \langle \hat{L}_j^\dagger \hat{L}_j \rangle\, \hat{\rho}$. Therefore, after taking the ensemble average, these two terms cancel with each other, and Eq.~\eqref{smeq2} reduces to the following Lindblad master equation:
\begin{equation}\label{lmeq2}
	\begin{aligned}
		\dot{\hat{\rho}} = -{i}[{\hat{H}},\hat{\rho}] + \sum_{j}\frac{\Gamma_{j}}{2}(2\hat{L}_j\hat{\rho}\hat{L}_j^{\dagger}-\hat{L}^{\dagger}_j\hat{L}_j\hat{\rho}-\hat{\rho}\hat{L}_j^{\dagger}\hat{L}_j).
	\end{aligned}
\end{equation}

Next, we consider a local observable consisting of a single term (single-term local observable) $\hat{O}_j$ associated with a set of mode $\{l\}_j$. The EOM of its mean value $O_j \equiv \langle\hat{O}_j\rangle$ is:
\begin{equation}\label{oeom}
	\begin{aligned}
		\dot{O}_j =&\, \text{Tr}\left[\dot{\hat{\rho}}\, \hat{O}_{j} \right]\\
		=&\, {i}\langle[{\hat{H}},\hat{O}_j]\rangle \\
                & + \sum_{j'} \frac{\Gamma_{j'}}{2} \left(2\langle\hat{L}_{j'}^{\dagger}\hat{O}_j\hat{L}_{j'}\rangle-\langle\hat{O}_j\hat{L}_{j'}^{\dagger}\hat{L}_{j'}\rangle-\langle\hat{L}^{\dagger}_{j'}\hat{L}_{j'}\hat{O}_j\rangle \right)\\
		=&\, {i}\langle[{\hat{H}},\hat{O}_j]\rangle \\
                & + \sum_{\{l\}_j} \frac{\Gamma_{l}}{2} \left(2\langle\hat{L}_{l}^{\dagger}\hat{O}_j\hat{L}_{l}\rangle-\langle\hat{O}_j\hat{L}_{l}^{\dagger}\hat{L}_{l}\rangle-\langle\hat{L}^{\dagger}_{l}\hat{L}_{l}\hat{O}_j\rangle \right),
	\end{aligned}
\end{equation}
where we have used the fact that $\hat{O}_j$ and $\hat{L}_{j'}$ with $j' \notin \{l\}_j$ commute with each other since $\hat{O}_j$ is an observable consisting of an even number of the creation/annihilation operators and both $\hat{O}_j$ and $\hat{L}_{j'}$ are single-term local quantities.
From the third line of Eq.~\eqref{oeom}, it is clear that, in the EOM of $O_j$, the contribution of the dissipator from the other modes $j' (\notin \{l\}_j)$ vanishes. Therefore, the jump terms from the other modes $j' (\notin \{l\}_j)$ do not cause any harm for the discussion.

If $\hat{O}_j$ is in normal order, the EOM (\ref{oeom}) can be further simplified.
Since we are considering a fermionic system with the Lindblad operator $\hat{L}_l$ for particle loss, the term $\langle\hat{L}_{l}^{\dagger}\hat{O}_j\hat{L}_{l}\rangle$ for normally ordered $\hat{O}_j$ vanishes. This occurs because $\hat{O}_j$ for $\{l\}_j$ contains at least one annihilation or creation operator for any mode in $\{l\}_j$, leading to a consecutive repetition of annihilation or creation operators for each mode $l$ in $\{l\}_j$.
It is noted that the vanishing of this term originated from the jump term is a natural consequence of focusing on the local quantity with a single term in dissipative fermionic systems. Resultingly, the EOM of ${O}_j$ becomes:
\begin{equation}\label{oeom2}
	\begin{aligned}
	  \dot{O}_j&={i}\langle[{\hat{H}},\hat{O}_j]\rangle - \sum_{\{l\}_j} \frac{\Gamma_{l}}{2} \left(\langle\hat{O}_j\hat{L}_{l}^{\dagger}\hat{L}_{l}\rangle+\langle\hat{L}^{\dagger}_{l}\hat{L}_{l}\hat{O}_j\rangle \right) \\
          &= i \langle \hat{H}^{\dagger}_{\text{eff}, j} \hat{O}_j - \hat{O}_j \hat{H}_{\text{eff}, j} \rangle
	\end{aligned}
\end{equation}
with
\begin{equation}\label{localnhh}
	\begin{aligned}
		\hat{H}_{\text{eff},j}=\hat{H}- i\sum_{\{l\}_j} \frac{\Gamma_l}{2} \hat{L}_l^\dagger \hat{L}_l. 
	\end{aligned}
\end{equation}
[These equations are identical to Eqs.~\eqref{meq} and \eqref{nhh2}, respectively.]

Note that there are no assumptions on $\hat{H}$ and the initial state in the above derivation, so that the local NHH formalism given by Eqs.~\eqref{nhh} and \eqref{meq} is applicable to arbitrary $\hat{H}$ and initial state.
As we have derived above, the local NHH gives the same EOM of $O_j$ as the one obtained from the Lindblad master equation. This also means that the local NHH is applicable to arbitrary timescales for which the original master equation is valid.

\begin{figure}[!tbp]
	\centering \includegraphics[width=1\linewidth]{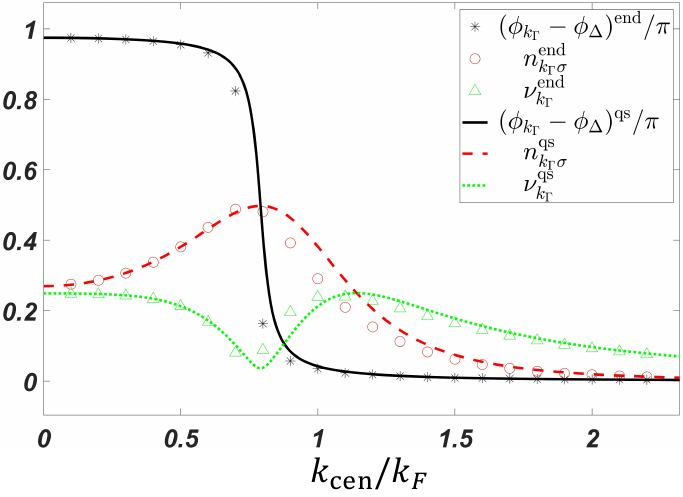} \caption{{\bf Comparison between numerical simulation and approximate analytical quasi-steady solution at $k_{{\Gamma}}$ in the small loss region.} The figure plots various quantities (relative phase, population, and anomalous average value) at $k_{\Gamma}$ in the quasi-steady state as functions of $k_{\text{cen}}$. The approximate analytical quasi-steady solutions given by Eq.\,\eqref{qst} with $k_{\Gamma}=k_{\text{cen}}$ are shown by lines: relative phase  $(\phi_{{k}_{{\Gamma}}}-\phi_{\Delta})^{\text{qs}}$ (black solid line), population $n_{k_{{\Gamma}}\sigma}^{\text{qs}}$ (red dashed line), and anomalous average value ${\nu}_{{k}_{{\Gamma}}}^{\text{qs}}$ (green dotted line). The results of the numerical simulation are shown by symbols: relative phase  $(\phi_{{k}_{{\Gamma}}}-\phi_{\Delta})^{\text{end}}$ (black star), population $n_{k_{{\Gamma}}\sigma}^{\text{end}}$ (red circle), and anomalous average value ${\nu}_{{k}_{{\Gamma}}}^{\text{end}}$ (green triangle). For the numerical results, we use the final state of the time evolution at $\omega_Ft=60$. The other parameters are same as Fig.\,\ref{FIG_pop}(c) (${1}/{k_F a_s}=0$, $k_c=10\,k_F$, $\delta_{{k}}=0.04\,k_F$, and $\Gamma =0.1\,\omega_F$).}
	\label{FIG_comparison}
\end{figure}

\bigskip
\noindent{\sf Verification of approximate analytical quasi-steady solution}\smallskip\\
We verify the approximate analytical quasi-steady solution given by Eq.~\eqref{qst}.

In Fig.\,\ref{FIG_comparison}, we compare this analytical quasi-steady solution with the corresponding results from the numerical simulation.
Here, we plot the relative phase $(\phi_{{k}_{{\Gamma}}}-\phi_{\Delta})$, population $n_{k_{{\Gamma}}\sigma}$, and anomalous average value ${\nu}_{{k}_{{\Gamma}}}$ in the loss region as the functions of the center of the loss region $k_{\text{cen}}$ both for the analytical (lines, with superscript ``qs'') and numerical (symbols, with superscript ``end'') results. For all of these quantities, the analytical results agree well with the numerical ones.
While Fig.\,\ref{FIG_comparison} shows the case of ${1}/{k_F a_s}=0$ with $\delta_{{k}}=0.04\,k_F$ and $\Gamma =0.1\,\omega_F$ as a concrete example, the qualitative results do not change throughout the BCS-BEC crossover as far as $\delta_k\,\Gamma$ is small enough so that $|\Delta|$ changes slowly over time.

\begin{figure}[!tbp]
	\centering \includegraphics[width=1\linewidth]{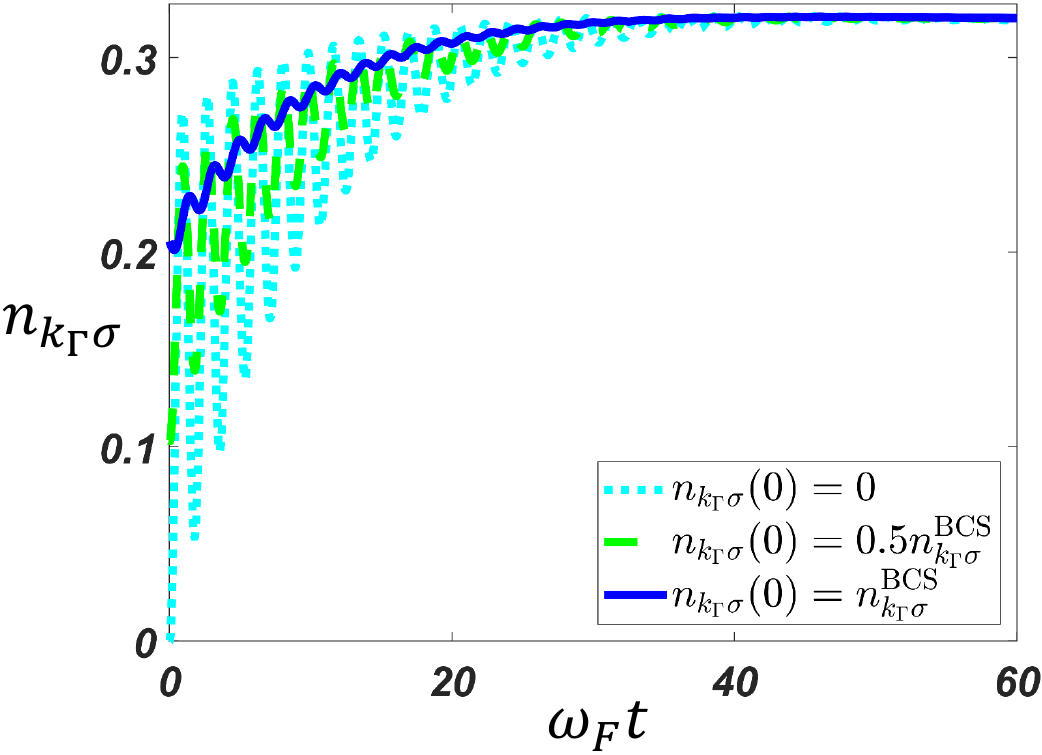}
	\caption{{\bf Robustness of the loss-induced population increase.} Time evolution of the population $n_{k_\Gamma\sigma}$ at $k_\Gamma$ in the small loss region for different initial states. The blue line is $n_{k_\Gamma\sigma}$ with the ground state $|\Psi_{\text{BCS}}\rangle$ as the initial state. The green dashed line is $n_{k_\Gamma\sigma}$ with $|\Psi_{\text{BCS}}\rangle$ except for $n_{k_\Gamma\sigma}=0.5n_{k_\Gamma\sigma}^\text{BCS}$ as the initial state. The cyan dotted line is $n_{k_\Gamma\sigma}$ with $|\Psi_{\text{BCS}}\rangle$ except for $n_{k_\Gamma\sigma}=0$ as the initial state. For clarity, we plot $n_{k_\Gamma\sigma}$ only for $k_{\text{cen}}$ in the loss region. The parameter values are the same as Fig.\,\ref{FIG_pop}(a) ($k_{\text{cen}}=0.4\,k_F$, ${1}/{k_F a_s}=1$, $k_c=15\,k_F$, $\delta_{{k}}=0.04\,k_F$, and $\Gamma =0.1\,\omega_F$).}
	\label{FIG_pop2}
\end{figure}

\bigskip
\noindent{\sf Robustness of the loss-induced population increase}\smallskip\\
Here, we demonstrate that even if starting from a different initial state other than the ground state $|\Psi_{\text{BCS}}\rangle$, the loss-induced population increase can also be obtained in a robust manner.

Since we are interested in the enhancement of the population $n_{k_\Gamma\sigma}$ at $k_\Gamma$ in the loss region, we focus on the population dynamics at $k_\Gamma$. To avoid violent oscillations by a strong quench, we prepare the initial states by changing the initial population only at $k_\Gamma$ from the population $n_{k\sigma}^\text{BCS}$ of the ground state $|\Psi_\text{BCS}\rangle$. In Fig.~\ref{FIG_pop2}, we show $n_{k_\Gamma\sigma}$ as a function of time for different initial states with $n_{k_\Gamma\sigma}(0) = 0$ (cyan dotted), $0.5 n_{k\sigma}^\text{BCS}$ (green dashed), and $n_{k\sigma}^\text{BCS}$ (blue solid). The other parameters are the same as in the case of Fig.\,\ref{FIG_pop}(a) ($k_{\text{cen}}=0.4\,k_F$, ${1}/{k_F a_s}=1$, $k_c=15\,k_F$, $\delta_{{k}}=0.04\,k_F$, and $\Gamma =0.1\,\omega_F$) [Thus, the blue solid line in Fig.~\ref{FIG_pop2} is the same as the blue solid line in Fig.~\ref{FIG_pop}(a).]. It is noted that, even if starting with $n_{k_\Gamma\sigma}$ smaller than $n_{k\sigma}^\text{BCS}$, all the cases converge to a same value larger than $n_{k\sigma}^\text{BCS} (\simeq 0.2)$ in the quasi-steady state. This means that, for these different initial states, we obtain almost the same quasi-steady state showing the loss-induced population increase. In addition, although not shown in the figure, the relative phase $\phi_{k_\Gamma} - \phi_\Delta$ of these cases is also locked at the same value as in the case of Fig.~\ref{FIG_pop}(a) in the quasi-steady state.

\bigskip
\noindent{\sf Spectral analysis of the local NHH for a simple model}\smallskip\\
Here, we perform a spectral analysis of the local NHH for a simple model to demonstrate that the spectrum of the local NHH carries the information of the dynamics of local observables. This information is hard to be identified in the spectra of the conventional NHH and the Liouvillian of the original Lindblad master equation.

We consider the following two-mode fermionic model:
\begin{align}
  \hat{H}=\frac{\Omega}{2}\left( \hat{c}_a^{\dagger} \hat{c}_b+\hat{c}_b^{\dagger} \hat{c}_a \right)-\frac{\omega}{2}\left( \hat{c}_a^{\dagger} \hat{c}_a-\hat{c}_b^{\dagger} \hat{c}_b \right),\label{eq:h2mode}
\end{align}
where $\hat{c}_l^\dagger$ ($\hat{c}_l$) is the creation (annihilation) operator of fermions in mode $l$ $(=a, b)$, $\omega$ and $\Omega$ are the energy difference and the Rabi frequency between the two modes, respectively. Then, we suppose that each mode is subject to one-body loss with the loss rate $\Gamma_l$ $(l=a, b)$ and the system is described by the Lindblad master equation given by
\begin{align}
  \dot{\hat{\rho}} = \mathcal{L} \hat{\rho} \equiv -i [\hat{H}, \hat{\rho}] + \sum_{l=a, b} \Gamma_l \mathcal{D}[\hat{c}_l] \hat{\rho}\label{eq:me2mode}
\end{align}
with the Liouvillian $\mathcal{L}$ and the dissipator $\mathcal{D}[\hat{c}_l] \hat{\rho} \equiv \frac{1}{2} (2 \hat{c}_l \hat{\rho} \hat{c}_l^\dagger - \hat{c}_l^\dagger \hat{c}_l \hat{\rho} - \hat{\rho} \hat{c}_l^\dagger \hat{c}_l)$.

This system has a closed set of coupled EOMs for the following three single-term local quantities, $\hat{O}_j$ $(j=1, 2, 3)$, defined for a set of modes $\{l\}_j$: $\hat{O}_1 \equiv \hat{n}_a \equiv \hat{c}_a^{\dagger} \hat{c}_a$ for $\{a\}_1$, $\hat{O}_2 \equiv \hat{n}_b \equiv \hat{c}_b^{\dagger} \hat{c}_b$ for $\{b\}_2$, and $\hat{O}_3 \equiv \hat{c}_a^{\dagger} \hat{c}_b = (\hat{c}_b^{\dagger} \hat{c}_a)^\dagger$ for $\{a, b\}_3$. These coupled EOMs read
\begin{align}
  \frac{\mathrm{d}\langle \hat{c}_a^{\dagger} \hat{c}_a\rangle}{\mathrm{d} t}=&\, -{i}\frac{\Omega}{2}\left(\langle \hat{c}_a^{\dagger} \hat{c}_b\rangle-\langle \hat{c}_b^{\dagger} \hat{c}_a\rangle\right)-\Gamma_a\langle \hat{c}_a^{\dagger} \hat{c}_a\rangle\,,\\
  \frac{\mathrm{d}\langle \hat{c}_b^{\dagger} \hat{c}_b\rangle}{\mathrm{d} t}=&\, -{i}\frac{\Omega}{2}\left(\langle \hat{c}_b^{\dagger} \hat{c}_a\rangle-\langle \hat{c}_a^{\dagger} \hat{c}_b\rangle\right)-\Gamma_b\langle \hat{c}_b^{\dagger} \hat{c}_b\rangle\,,\\
  \frac{\mathrm{d}\langle \hat{c}_a^{\dagger} \hat{c}_b\rangle}{\mathrm{d} t}
  =&\,-{i}\frac{\Omega}{2}\left(\langle \hat{c}_a^{\dagger} \hat{c}_a\rangle-\langle \hat{c}_b^{\dagger} \hat{c}_b\rangle\right)-{i}\omega\langle \hat{c}_a^{\dagger} \hat{c}_b\rangle\nonumber\\
  &\, -\frac{\Gamma_a+\Gamma_b}{2}\langle \hat{c}_a^{\dagger} \hat{c}_b\rangle\,,\\
  \frac{\mathrm{d}\langle \hat{c}_b^{\dagger} \hat{c}_a\rangle}{\mathrm{d} t}=&\, \left(\frac{\mathrm{d}\langle \hat{c}_a^{\dagger} \hat{c}_b\rangle}{\mathrm{d} t}\right)^*\,.
\end{align}

The local NHHs $\hat{H}_{\text{eff},j}$ for $\hat{O}_j$ with $j=1$ and $2$, i.e., $\hat{O}_1 = \hat{c}_a^{\dagger} \hat{c}_a$ and $\hat{O}_2 = \hat{c}_b^{\dagger} \hat{c}_b$, are
\begin{align}
  \hat{H}_{\text{eff}, j} = \hat{H} - i \frac{\Gamma_{\{l\}_j}}{2} \hat{c}_{\{l\}_j}^\dagger \hat{c}_{\{l\}_j}\qquad (j=1, 2)\,\label{eq:lnhh2mode}
\end{align}
with $\{l\}_1 = a$ and $\{l\}_2 = b$.
The local NHH $\hat{H}_{\text{eff},3}$ for $\hat{O}_3 = \hat{c}_a^{\dagger} \hat{c}_b$ is
\begin{align}
  \hat{H}_{\text{eff}, 3} = \hat{H} - i \sum_{l=a,b} \frac{\Gamma_l}{2} \hat{c}_l^\dagger \hat{c}_l\,.\label{eq:lnhh2mode2}
\end{align}
The conventional NHH $\hat{H}_{\text{eff}}$ obtained by neglecting the jump terms of the Lindblad master equation (\ref{eq:me2mode}) is given by
\begin{equation}
  \hat{H}_{\text{eff}} = \hat{H} - i \sum_{l=a, b} \frac{\Gamma_l}{2} \hat{c}_l^\dagger \hat{c}_l\,,\label{eq:nhh2mode}
\end{equation}
which is identical to one of the local NHHs, $\hat{H}_{\text{eff},3}$, corresponding to $\hat{O}_3$, as given by Eq.~(\ref{eq:lnhh2mode2}).

\begin{figure}[!tbp]
	\centering \includegraphics[width=1\linewidth]{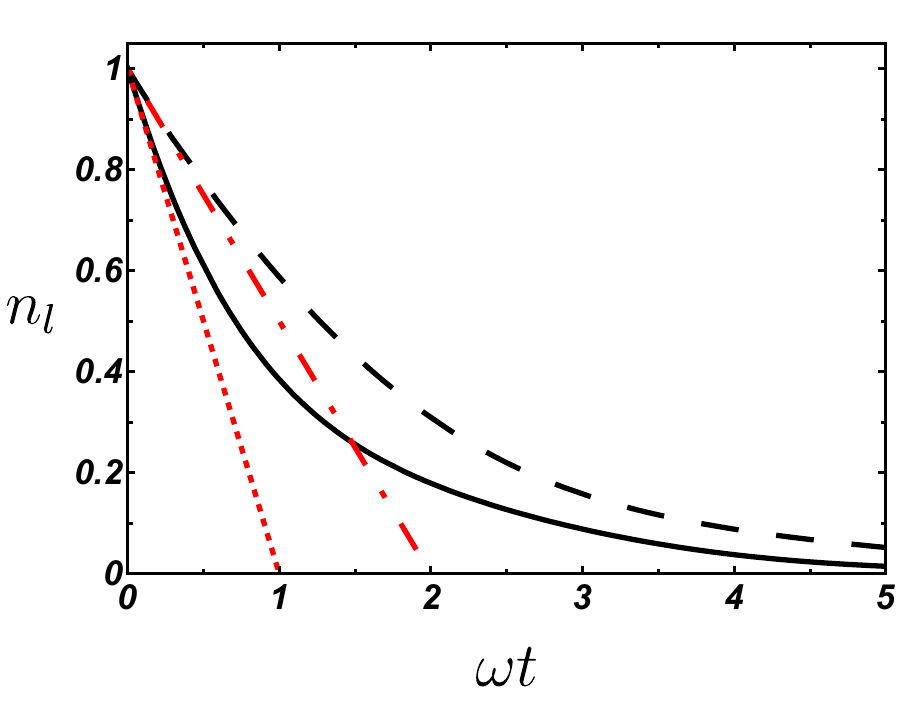}
	\caption{{\bf Population dynamics and its time constant.} Time evolution of the population $n_l$ of mode $l=a$ and $b$ obtained by solving the Lindblad master equation (\ref{eq:me2mode}). Here, the population dynamics (black lines) of each mode is compared with the time constant (red lines) predicted by the eigenvalues of the local NHHs $\hat{H}_{\text{eff}, j}$ $(j=1, 2)$: population $n_a$ (black solid line) and $n_b$ (black dashed line), time constant $\Gamma_a^{-1}$ (red dotted line) and $\Gamma_b^{-1}$ (red dash-dotted line). The initial state is $\hat{\rho} = |1,1\rangle \langle 1,1|$ with $|n_a, n_b\rangle$ being the Fock state. The parameter values are $\Omega = 1$, $\Gamma_a = 1$, and $\Gamma_b = 0.5$ in units of $\omega$.}\label{FIG_pop2mode}
\end{figure}

Now, let's consider the population dynamics of mode $a$ and $b$.
The eigenvalues $\{E^{(j)}_n\}$ of the local NHH $\hat{H}_{\text{eff},j}$ for $j=1$ and $2$ corresponding to $\hat{O}_1=\hat{n}_a$ and $\hat{O}_2=\hat{n}_b$, respectively, are
\begin{align}
  \{E^{(j)}_n\} =&\, \bigg\{0, -i \frac{\Gamma_{\{l\}_j}}{2}, \nonumber\\
  &\quad - \frac{i}{4}\left[\Gamma_{\{l\}_j} \pm \sqrt{(\Gamma_{\{l\}_j} - 2i\omega)^2 - 4\Omega^2}\,\right] \bigg\}\,.\label{eq:lnhhspectrum2mode}
\end{align}
In Fig.~\ref{FIG_pop2mode}, we plot the time evolution of the population $n_l \equiv \text{Tr}[\hat{\rho}\, \hat{n}_l]$ of each mode $l=a$ and $b$ calculated from the Lindblad master equation (\ref{eq:me2mode}). Here we note that the population $n_l$ decays with a time constant $\Gamma_l^{-1}$ for each mode $l$ separately as predicted from the spectrum (\ref{eq:lnhhspectrum2mode}) of each local NHH $\hat{H}_{\text{eff},j}$ for $j=1$ and $2$.

On the other hand, $\Gamma_a$ and $\Gamma_b$ always appear in pair as $\Gamma_a \pm \Gamma_b$ in all the eigenvalues of the Liouvillian $\mathcal{L}$ and the conventional NHH $\hat{H}_{\text{eff}}$ (thus, it is also the case for the local NHH $\hat{H}_{\text{eff},3}$ for $\hat{O}_3=\hat{c}_a^\dagger \hat{c}_b$). For example, the eigenvalues $\{E_n\}$ of the conventional NHH $\hat{H}_{\text{eff}}$ (same for $\hat{H}_{\text{eff},3}$) are
\begin{align}
  \{ E_n \} =& \bigg\{0, -\frac{i}{2} (\Gamma_a + \Gamma_b),\nonumber\\
  & -\frac{i}{4} \left[\Gamma_a + \Gamma_b \pm \sqrt{(\Gamma_a - \Gamma_b - 2i \omega)^2 - 4 \Omega^2} \right] \bigg\}\,.
\end{align}
It is noted that the time constant $\Gamma_a^{-1}$ or $\Gamma_b^{-1}$ cannot be obtained from any combination of these eigenvalues.

\bigskip
\noindent{\bf ACKNOWLEDGMENTS}\smallskip\\
We thank Chao Gao for helpful discussions and comments. G.\,W. acknowledges support from the National Natural Science Foundation of China (Grant No\,12375039 and No.\,11975199), the Zhejiang Provincial Natural Science Foundation Key Project (Grant No.\,LZ19A050001), and the Zhejiang University 100 Plan.


\bibliography{references_submit3}

\end{document}